\documentclass[runningheads,12pt]{llncs}
\usepackage{graphicx}

\usepackage[utf8]{inputenc} 
\usepackage[T1]{fontenc}    
\usepackage{hyperref}       
\usepackage{url}            
\usepackage{booktabs}       
\usepackage{amsfonts}       
\usepackage{nicefrac}       
\usepackage{microtype}      
\usepackage{lipsum}
\usepackage{color}
\usepackage{verbatim}
\usepackage{a4wide}
\usepackage{comment}

\makeatletter

\newcommand{\Rmnum}[1]{\expandafter\@slowromancap\romannumeral #1@}
\makeatother

\usepackage[
	n,
	operators,
	advantage,
	sets,
	adversary,
	landau,
	probability,
	notions,	
	logic,
	ff,
	mm,
	primitives,
	events,
	complexity,
	asymptotics,
	keys]{cryptocode}

\usepackage{dashbox}
\usepackage{tikz}

\usepackage{pgfplots}
\definecolor{bblue}{HTML}{4F81BD}
\definecolor{rred}{HTML}{C0504D}
\definecolor{ggreen}{HTML}{9BBB59}
\definecolor{ppurple}{HTML}{9F4C7C}
\definecolor{pgolden}{HTML}{8B814C}

\usepackage{listings} 
\usepackage{color} 
\definecolor{mygreen}{rgb}{0,0.6,0}
\definecolor{mygray}{rgb}{0.5,0.5,0.5}
\definecolor{mymauve}{rgb}{0.58,0,0.82}
\definecolor{darkgray}{rgb}{.4,.4,.4}
\definecolor{purple}{rgb}{0.65, 0.12, 0.82}

\renewcommand\rightmark{Auditable Credential Anonymity Revocation}
\renewcommand\leftmark{R. Li, D. Galindo, and Q. Wang}
\newcommand{\romannum}[1]{\romannumeral #1}

\begin{document}
\title{Auditable Credential Anonymity Revocation Based on Privacy-Preserving Smart Contracts}


 \author{Rujia Li\inst{1,2} \and
David Galindo \inst{2,3} \and
Qi Wang\inst{1}}
 

\institute{Southern University of Science and Technology, Shenzhen, China \and
University of Birmingham, United Kingdom \\ \and
Fetch.AI, Cambridge, United Kingdom\\ \medskip
\email{\{rxl635,d.galindo\}@cs.bham.ac.uk}\\
\email{wangqi@sustech.edu.cn}
}


\maketitle

\begin{abstract}

Anonymity revocation is an essential component of credential issuing systems since unconditional anonymity is incompatible with pursuing and sanctioning credential misuse. However, current anonymity revocation approaches have shortcomings with respect
to the auditability of the revocation process. In this paper, we propose a novel anonymity revocation approach based on privacy-preserving blockchain-based smart contracts, where the code self-execution property ensures availability and public ledger immutability provides auditability.  We describe an instantiation of this approach, provide
an implementation thereof and conduct a series
of evaluations in terms of running time, gas cost and latency. The results show that our scheme is feasible and efficient.


\end{abstract}

\section{Introduction}

Anonymity revocation was first discussed by von Solms and Naccache~\cite{von_solms_blind_1992}, as they pointed out that Chaum's blind signatures~\cite{chaum_blind_1983} could potentially lead to nonpunishable crime. Subsequently, anonymity revocation has been studied comprehensively, especially in e-cash systems designed to combat money laundering and blackmailing~\cite{blazy2011achieving,camenisch_digital_1996,canard2003fair}.  The idea of adding anonymity revocation to anonymous credential systems was first proposed by Camenisch and Lysyanskaya~\cite{camenisch_efficient_2001}, where they offered an optional anonymity tracing approach to find the identity of pseudonymous tokens involved in suspicious transactions. In general, anonymity revocation in a credential system allows an issuer to find out 
who the owner of an anonymous credential is.

The blindness issuance property of an anonymous credential system prevents an issuer from completing the task of anonymity revocation by themselves. The party who helps the issuer to reveal the identity is referred to as {\em revelator}. Intuitively, there are two parties that can act as the revelator: the user (credential holder) and the judge (trusted third party). Voluntary anonymity revocation by the user is usually straightforward. The issuer cannot link the identity, the message and the resulting signature together unless the user does. One typical example is Microsoft's U-Prove~\cite{paquin_u-prove_2011}.  In such a system, the issuing protocol and the showing protocol are unlinkable. Even if the issuer colludes with the verifier, it cannot associate the message with the resulting signature. The only possibility is that the user chooses  to lift the anonymity. Meanwhile, lifting anonymity by a judge, which is inspired by fair blind signature scheme~\cite{fuchsbauer_fair_2010}, is widely used in systems such as~\cite{park_traceable_2009,camenisch_formal_2010,escala_revocable_2011,rannenberg_attribute-based_2015}. Taking ABC4Trust~\cite{rannenberg_attribute-based_2015} as an example, it introduced an inspector to uncover the user who created a presentation token to prevent abuse.

However, some weaknesses in the mentioned anonymity revocation approaches still remain. Firstly, revealing anonymity through the credential holder relies too heavily on the user's will, which ultimately leads to the nonavailability  problem. This means if a user behaves maliciously and rejects to cooperate with the issuer, the issuer would never learn the relationship between the identity and the credential. Furthermore, even if the user is honest, they may be offline, resulting in the failure of blindness removal. Meanwhile, in the majority of previous proposals revealing anonymity through the judge lacks transparency, which raises some security concerns: (1) even without the user’s consent, the issuer and the judge may conspire to map the credential to the real identity of that user; (2) the judge is a single point of failure. More importantly, the user has no auxiliary information to detect whether the judge has been compromised or not. These challenges lead to the following question:

\begin{quote}
\textit{Is it possible to build an anonymity revocation mechanism that satisfies the requirements: (1) the process of lifting anonymity is transparent and auditable; (2) the revelator always accept revealing the anonymity if necessary?}
\end{quote}

In this paper, we give a positive answer to this question. Instead of using a trusted third party, we use a neutral and transparent privacy-preserving smart contract as the revelator (to revoke the blindness). The self-execution property of the smart contract ensures the availability of the revelator. This means the neutral blockchain is always honest and is willing to revoke the anonymity whenever it is needed by the issuer. Meanwhile, our privacy-preserving smart contract-based approach allows anonymity revocation in an auditable manner. More precisely, the anonymity tracing must interact with the privacy-preserving smart contract that ``lives'' on the blockchain and automatically renders the progress auditable. Such revocation progress is recorded in a blockchain transaction which is publicly visible. This auditability provided by the smart contract calling records avoids the misuse of revocation and reduces potential collusion problems to a great extent. Furthermore, the transparent contract calling records provide the user with auxiliary information to detect whether the issuer has been compromised.

In addition, our scheme brings the benefit of greater availability. The service of blockchain is maintained by a large group of nodes~\cite{nakamoto_bitcoin:_2008}, which avoids the offline revelator problem. Alternatively, the high-availability blockchain service, being continuously online, provides greater actualization of blindness disclosure and the tracer could trace the identity or credential at any time. \medskip

In summary, the contributions of this paper can be summarized as follows: 

\begin{itemize}

\item We propose a new auditable blind credential system based on privacy-preserving smart contracts, which provides a powerful auditability and neutrality for credential anonymity revocation

\item We give an instantiation of our construction and provide a proof of concept implementation. The performance evaluation shows that our scheme is feasible and efficient.

\end{itemize}

The rest of the paper is structured as follows: further related work is discussed in Section~\ref{sec:work}. Notation and cryptographic building blocks are presented in Section~\ref{sec:preliminaries}. An overview of our construction is given in Section~\ref{sec:overview}, followed by an instantiation in Section~\ref{sec:instantiation}. The implementation and evaluation of the instantiation are detailed out in Section~\ref{sec:implementation}. Some example applications are given in Section~\ref{sec:application}. Finally, Section~\ref{sec:conclusion} concludes with some future work.

\section{Related Work}
\label{sec:work}

In this section, we first survey current anonymity revocation approaches and make a comparison with our solution. Then, we give some background on blockchain and  privacy-preserving smart contracts.

In the last few decades, a series of works~\cite{stadler_fair_1995,abe_provably_2001,kiayias_concurrent_2006,fuchsbauer_fair_2010}, have been proposed in the field of anonymity revocation, especially in e-cash systems. Brickell \textit{et al.}~\cite{Brickell1995TrusteebasedTE} introduced the first trustee-based tracing electronic cash system, in which the coin owner can be revealed by several publicly appointed trustees. Camenisch \textit{et al.} ~\cite{camenisch_digital_1996} proposed an anonymous digital payment system with a passive anonymity-revoking trustee. In their system, the trustee only needs to be involved in the anonymity-revoking progress rather than the regular transactions such as opening a new account. Jakobsson and Yung~\cite{Jakobsson1996RevokableAV} presented an e-money system that makes the value of funds and user anonymity revocable with the consumer rights organisations, even given an extreme condition that an active attacker gets the bank's key or forces the bank to release the money.

In 1995, a fair blind signature scheme was first proposed by Stadler \textit{et al.}~\cite{stadler_fair_1995}. It involved a judge and allowed this judge to deliver information to the signer to link the issuing session and the resulting message-signature pair. Later, Jakobsson and Yung~\cite{jakobsson_distributed_1997} pointed out that the reused session identifier may make the anonymity revocation invalid, and proposed a fair blind signature scheme that guarantees the one-to-one mapping revocability between the issuing session and the resulting signature. Thereafter, Hufschmitt \textit{et al.}~\cite{hufschmitt_fair_2007} presented a formal security model for fair blind signatures in the random oracle model. Then, based on Hufschmitt's model, Fuchsbauer \textit{et al.}~\cite{fuchsbauer_fair_2010} proposed a fair blind signature scheme that is not based on the random oracle model. To the best of our knowledge, Camenisch and Lysyanskaya ~\cite{camenisch_efficient_2001} was the first to use anonymity revocation in the credential system. They offered an optional approach to trace the identity of the pseudonymous token for some transactions. After that, some practical systems like IBM’s Identity Mixer~\cite{camenisch_formal_2010}, ABC4Trust~\cite{rannenberg_attribute-based_2015} started to consider the anonymity revocation. An interesting revocation approach is traceable anonymous certificate~\cite{kwon_privacy_2011}. It allows one sub-issuer to verify the ownership of a user and another sub-issuer to validate the contents. Then, these two issuers collaborate to map the certificate to its real identity.

However, the aforementioned anonymity revocation approaches have some drawbacks: the repudiation and the lack of auditability in the revocation progress. The assumptions that the revelator always remains honest is unrealistic. The revelator may be offline when it is needed, or may conspire with the issuer to seek profits, or even be entirely controlled by an attacker. Our scheme is the first to use a privacy-preserving smart contract as the revelator to solve the above problems. The self-executing nature of the contract ensures the neutrality of the revelator. The transparent contract calling records guarantee that the revelator's revocation progress is auditable. The continuous blockchain service keeps the high-availability of the revelator.

\medskip

\noindent \textbf{Privacy-preserving smart contracts.} The concept of a \textit{smart contract}, as a primary application of blockchain technologies~\cite{nakamoto_bitcoin:_2008}, was first proposed by Szabo~\cite{szabo1996smart}. It is originally defined as a set of digital protocols within which the parties abide by some pre-agreed commitments. In the blockchain system, the smart contract is designed as a self-executing protocol that can verify or execute the fulfilment for the shared instruction code. The smart contract is generally made up of two parts: the instruction code and the executed status. The smart contract in the traditional blockchain systems such as Bitcoin and Ethereum~\cite{luu_making_2016} lacks privacy since the instruction code and executed status are publicly shared and visible among all the participants (nodes) in the network. Recently, a new line of work~\cite{kosba_hawk:_2016,mccorry_smart_2017,kalodner_arbitrum:_2018,bunz_zether:_2019} claimed that they had solved these privacy issues by proposing the privacy-preserving smart contract platforms. To verify the feasibility of our scheme,  we selected Ekiden~\cite{cheng_ekiden:_2018} and its implementation Oasis Devnet as our privacy-preserving platform. Ekiden~\cite{cheng_ekiden:_2018} combines trusted execution environments (TEEs) and blockchain to achieve confidentiality as well as decentralisation. It allows replicating the contract execution to TEE-powered nodes, where these TEE-powered nodes guarantee the private state and data of the contracts by encrypting them with crytographic keys only known to them.

\section{Preliminaries}
\label{sec:preliminaries}

In this section, we define the notation and recall a well-known cryptographic building block that will be used in our construction.

\subsection{Notation}
Let $\lambda$ be the security parameter,  $\Sigma(\mathsf{KeyGen},\mathsf{Sig},\mathsf{Vf})$ represent a standard signature scheme, and $\mathcal{SM}.\mathsf{Enc}(\mathsf{KeyGen},\mathsf{Enc},\mathsf{Dec})$ stand for  symmetric encryption and $\mathcal{ASM}.\mathsf{Enc}(\mathsf{KeyGen},\mathsf{Enc},\allowbreak\mathsf{Dec})$ refer to  asymmetric encryption.

\subsection{Fair blind signature}
Informally, a fair blind signature is an interactive protocol between three parties: the user, the issuer and the tracer. It is defined by eight probabilistic polynomial-time algorithms $\mathsf{Setup}$, $\mathsf{KeyGen}$,  $\mathsf{Issue_{sig}}$, $\mathsf{Verify_{sig}}$,$\mathsf{Trace_{sig}}$, $\mathsf{Trace_{id}}$, $\mathsf{Match_{sig}}$, and $\mathsf{Match_{id}}$ as follows. For a formal functional definition of fair blind signature schemes see for example ~\cite{abe_provably_2001},~\cite{fuchsbauer_fair_2010}.

\begin{itemize}

\item $\mathsf{Setup}$ is a parameter generation algorithm that takes the security parameter $\lambda$, and outputs the common parameters $params$ for the following algorithms;  $params \gets \mathsf{Setup} (1^{\lambda})$. 

\item $\mathsf{KeyGen}$ is a key generation algorithm that takes the parameter $params$, and outputs a key pair $(\sk, \pk)$; $(\sk, \pk) \gets \mathsf{KeyGen} (params)$.

\item $\mathsf{Issue_{sig}}$ is an algorithm that takes the message $msg$ and outputs a blind signature; $\Sigma_{mgs} \gets \mathsf{Issue_{sig}} (msg)$.

\item $\mathsf{Verify_{sig}}$ is an algorithm to verify the signature $\Sigma_{mgs}$. It outputs $1$ if $sig_{m}$ is valid, and $0$ otherwise; $0/1 \gets \mathsf{Verify_{sig}}(\Sigma_{mgs})$.

\item $\mathsf{Trace_{sig}}$ is a revocation algorithm that generates a resulting signature $sig_{m}^{'}$, where this signature is yielded from the target session identifier $id_{u}$; $sig_{u}^{'} \gets \mathsf{Trace_{cred}}(id_{u})$.

\item $\mathsf{Trace_{id}}$ is a revocation algorithm that generates the session identifier $id_{u}^{'}$ which has produced target signature $sig_{u}$; $id_{u}^{'} \gets \mathsf{Trace_{id}}(sig_{u})$.

\item $\mathsf{Match_{sig}}$ is a matching algorithm that examines whether the original signature $sig_{u}$ matches to the resulting signature $sig_{u}^{'}$ or not. It outputs $1$ if they match, and $0$ otherwise; $0/1 \gets \mathsf{Match_{sig}}(sig_{u}, sig_{u}^{'})$.

\item $\mathsf{Match_{id}}$ is a matching algorithm that examines whether  the original session identifier $id_{u}$ matches to the resulting identifier $id_{u}^{'}$ or not. It outputs $1$ if they match, and $0$ otherwise; $0/1 \gets \mathsf{Match_{id}}(id_{u}, id_{u}^{'})$.
\end{itemize}

\section{Construction Overview}
\label{sec:overview}

\begin{figure}[htb!]
    \centering
    \includegraphics[width=11cm]{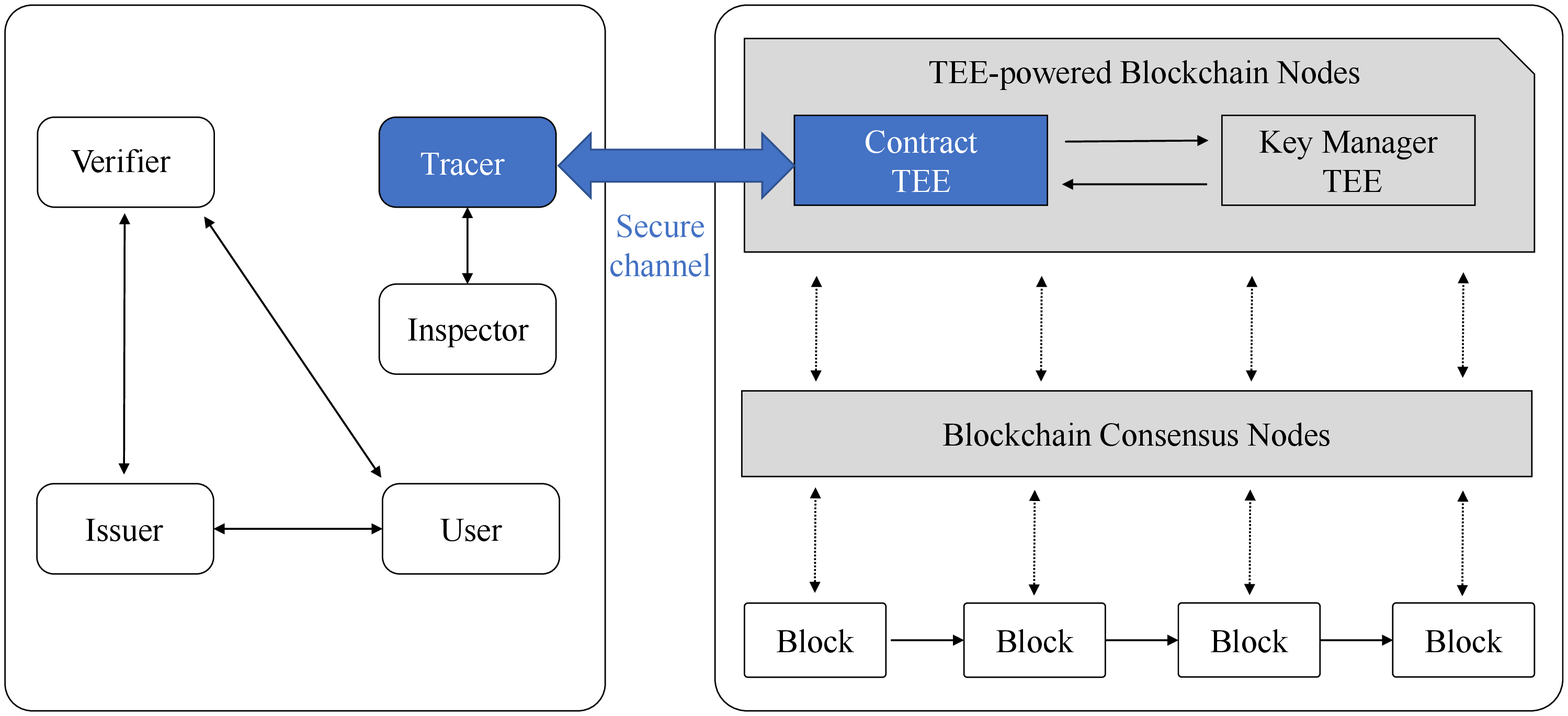}
    \caption{Overview of our construction}
    \label{fig:construction}
\end{figure}

An auditable blind credential system has six participants (see Fig.~\ref{fig:construction}): the issuer, the user, the verifier, the tracer, the inspector and the privacy-preserving smart contract platform. The user is the holder of a credential. The issuer is in charge of blindly issuing a credential. The verifier is responsible for checking the validity of the credential. The tracer is used to reveal the relationship of the credential and its identity. It is noted that, to have a clear understanding, we introduce the concept of tracer and allow both the issuer and the verifier to act as the tracer. The inspector is used to check the suspicious revocation activities and report them. The privacy-preserving smart contract platform is employed as a revelator to provide the revocation service. The privacy-preserving smart contract platform includes two types of blockchain nodes: the TEE-powered blockchain nodes and the consensus nodes. The TEE-powered blockchain nodes are composed of the contract TEE and the key manager TEE, where contract TEE is used to execute the smart contract and then encrypt the resulting state with the key from the key manager TEE. The consensus nodes are used to achieve the agreement of the encrypted state of the smart contract.

\begin{table}[htb!]
\caption{A high-level description of anonymity revocation with blockchain}
\label{table:2}
\centering
\begin{tabular}{p{\textwidth}}
\hline \\
\textbf{System Setup}\\
$params \gets \mathsf{Setup} (1^{\lambda})$; the system takes $1^{\lambda}$ and outputs the system parameters $params$.
$(\sk_{*}, \pk_{*}) \gets \mathsf{KeyGen_{entities}} (params)$; the entities (issuer, user, tracer) input $params$ and output their key pair $(\sk_{*}, \pk_{*})$.\\ \\

 \textbf{Smart Contract Registration} \\
$\widehat{contract} \gets \mathsf{Deploy_{contract}}(params, code)$; the system takes $params$ and a piece of contract code $code$ and outputs the privacy-preserving smart contract $\widehat{contract}$. \\
$(\sk_t, \pk_t) \gets \mathsf{KeyGen_{ppsc}} (params,\widehat{contract})$; given $params$ and $\widehat{contract}$, $\mathsf{KeyGen_{ppsc}}$ generates the tracing key paris $(\sk_t, \pk_t)$. \\ \\

\textbf{Credential Generation} \\
$sig_{attrs} \gets \mathsf{Issue_{sig}} (attrs, \pk_t, \dots)$; the issuer inputs the user's attributes $attrs$ and public tracing key $\pk_t$, etc., and outputs the signature of these attributes. \\
$cred_{u} \gets \mathsf{FormCred} (attrs, sig_{attrs})$; the issuer inputs the attributes and its signature and outputs a credential. \\ \\

\textbf{Credential Verification} \\
$0/1 \gets \mathsf{Verify_{sig}} (cred_u)$; the verifier checks the signature of the credential $cred_u$ with output $0$ or $1$. \\ \\

\textbf{Credential Tracing} \\ 
$cred_{u} \gets \mathsf{Trace_{cred}}(id_{u}^{'})$; $\mathsf{Trace_{cred}}$ takes the identity $id_{u}^{'}$ and outputs the credential of that identity. \\
$tran_{cred} \gets \mathsf{FormTrans}(cred_{u},\widehat{contract})$; the tracer invokes $\widehat{contract}$ to obtain the $tran_{cred}$ that contains the encrypted $cred_{u}$. \\ \\

\textbf{Identity Tracing} \\
$id_{u} \gets \mathsf{Trace_{identity}}(cred_{u}^{'})$; $\mathsf{Trace_{identity}}$ takes the credential $cred_{u}^{'}$ and outputs the identity of that credential. \\
$tran_{id} \gets \mathsf{FormTrans}(id_{u},\widehat{contract})$; the tracer invokes  $\widehat{contract}$ to obtain the $tran_{id}$ that contains the encrypted $id_{u}$. \\ \\

\textbf{Tracing Inspection} \\
$views_{t} \gets \mathsf{Collect_{trans}}(\pk_t, type)$; the inspector scans the blockchain to collect the tracer's invoking records (represented as the transactions) depending on the type of identity tracing or credential tracing. \\
$0/1 \gets \mathsf{Inspect_{trans}}(views_{t})$; the inspector takes the $views_{t}$ and outputs the inspection result.\\ \\
\hline
\end{tabular}
\end{table}

In general, a basic version of auditable blind credential system works as follows: the system sets up the parameters and prepares for the key pairs for the issuer, the user and the tracer. Then, the system sends a smart contract to a TEE-powered blockchain node to obtain a privacy-preserving contract $\widehat{contract}$, in which the method name, arguments, and return data are externally invisible. Then, the system invokes the $\widehat{contract}$ through contract TEE to generate the tracing key pair $(x_t,y_t)$. The private key $x_t$ is kept secretly, and only contract TEE can access it internally. Key $y_t$ is public and is used in the issuing protocol. Next, the user authenticates himself to the issuer to obtain an anonymous credential. After that, the user shows the credential to the verifier who wants to check the validity. So far, due to the blind issuance, neither the issuer nor the verifier knows the relationship of the credential and its holder. 

In the revocation stage, the tracer firstly builds an encrypted and authenticated channel with the contract TEE (one crucial property of remote attestation~\cite{mckeen_innovative_2013} in TEE). Then, given the user's identifier or the anonymous credential, the $\widehat{contract}$ lifts the blindness and returns the result to the tracer bearing a transaction. Due to the protection of the encrypted channel, the contents of the transaction including the input and the output data are kept secret. However, the invoking records of the transaction remain visible and become immutable because of the confirmation by the consensus nodes. Alternatively, any entity can see the fact that the tracer is interacting with the contract, but nobody except the tracer knows the exact data in the transaction. Subsequently, the inspector scans the blockchain to collect the tracer’s calling records and inspect the suspicious credential tracing activity. We give a high-level description in Table~\ref{table:2}.

\section{Concrete Instantiation} 
\label{sec:instantiation}

In this section, we present an instantiation based on Abe's~\cite{abe_provably_2001} blind signature scheme and the privacy-preserving smart contract platform of Ekiden~\cite{cheng_ekiden:_2018}. For security and efficiency purposes, we slightly modified Abe's ~\cite{abe_provably_2001} scheme by using elliptic curve cryptography. Thus, all the following arithmetic operations are based on addition of points and hereafter unless otherwise noted. 

Let $\mathcal{G}$ be a probabilistic polynomial-time algorithm that generates an elliptic curve group $(\mathrm{E}(\ZZ_p), p,q, g, h) \gets \mathcal{G}(1^{\lambda})$, where $p$ is a big prime number, $q$ is the order and $(g, h)$ are elements of $\mathrm{E}(\ZZ_p)$. Hash functions $\mathcal{H}_{1} : \{0, 1\}^{\star} \to \langle g \rangle$,and $\mathcal{H}_{2}, \mathcal{H}_{3}: \{0, 1\}^{\star} \to  \{0, 1\}^{|q|}$ are defined.
The function $\mathcal{H}_{1}$ refers to mapping an arbitrary string to an element of the subgroup $\langle g \rangle$ and function $\mathcal{H}_{2}$ and function $\mathcal{H}_{3}$ all refer to mapping an arbitrary string to an element of $\ZZ_q$ with the fixed length.

\paragraph{Key Generation} The issuer generates a  public key $y$ and a tag key $z$, where $ x \in \ZZ_q\; , \; y = g^x \, mod \, q $ and $z = \mathcal{H}_{1}(p, q, g, h, y) $. A user generates a key pair $(\gamma,\xi)$, where $\gamma \in \ZZ_q\; $ and $  \xi = g^{\gamma} \, mod \, q\;$. To simplify the instantiation, we use the session identifiers to represent the user's identity and allow one user to generate multiple identities $(\gamma_1, \xi_1), (\gamma_2, \xi_2), \dots (\gamma_n, \xi_n)$. Similarly, the tracer generates the session key pair $(\iota,\tau)$, where $\iota \in \ZZ_q\;$ and $\tau = g^{\tau} \, mod \, q$. It should be noted that the tracer's session key is only used to establish the authenticated channels to the contract TEE.

\paragraph{Contract Registration} The system compiles pieces of code of a smart contract $\widehat{contract}$ and sends its bytecode to a TEE-powered blockchain node. Then, the TEE-powered blockchain node first loads bytecode into the contract TEE. Then, the contract TEE creates a new contract identifier $ppsc$, obtains a fresh internal contract key pair $(\pk_{cid}^{in},\sk_{cid}^{in})$ and an internal state key $\key_{state}$ from the key manager TEE. Thereafter, the contract TEE outputs an encrypted initial contract state $state_{init} = \mathcal{SM}.\mathsf{Enc}(\key_{state}, \overrightarrow{state_0})$ and an attestation $\Omega_{cid}$, where $\Omega_{cid}$ is used to prove the correctness of this initialization. After that, the TEE-powered blockchain node gets a proof $\pi$ of $\Omega_{cid}$ by the attestation service and push the final composition $(\widehat{contract},\pk_{cid}^{in},state_{init},\Omega_{cid},\pi)$ to the blockchain consensus nodes. The blockchain consensus nodes would like to accept this smart contract if all the attestations and proofs are verified successfully. As for parameter registration, given the common parameters of $\mathrm{E}(\ZZ_p)$ and the public key of an issuer, say $\pk_i$, $\widehat{contract}$ takes a random number $x_t$ under $\ZZ_q$ as the private tracing key and generates its public key $ y_t = g^{x_t} \, mod \, q $. The private key $x_t$ is held by the secret state, which can only be accessed by the contract TEE internally.

\subsection{Blind Issuance} 

\paragraph{Credential Generation} Credential generation is an interactive protocol that involves only the user and issuer, which means that it runs independently from the privacy-preserving smart contract. The main idea of this protocol is witness indistinguishable~\cite{Feige1990WitnessIA}. Namely, the issuer owns a key pair $(x, y)$ where $x \in \ZZ_q, y = g^x \, mod \, q\;$, and a ``one-time'' tag key pair $(\omega, z)$ , where $\omega \in \ZZ_q, z = g^{\omega}$. The signature can only be issued by the real private key $x$ but no one can distinguish which of the two secret keys  ($x$ or $\omega$) was used. A full description is as follows: The user firstly computes $z_u = z^{1/\gamma}$ and proves to the issuer that $\log_g \xi$ is equal to $\log_{z_u} z$. Then, the issuer generates random string $\upsilon$, and computes $z_1 = y_t^{\upsilon}$ and $z_2 = z_u / z_1$, and then proves to the user that $z_1$ is made as it should be. Based on $y$, $z_1$, $z_2$, the issuer and the user engage in an interactive proof protocol, in a witness indistinguishable way, to prove the knowledge of the following two parts:

\begin{itemize}
    	\item \textbf{y-side:} proof of knowledge of $x$ of $y = g^x$.
        \item \textbf{z-side:} proof of knowledge of $(\omega_1, \omega_2)$ of $b_1 = g^{\omega_1},b_2 = g^{\omega_2}$.
\end{itemize}

After that, the user blinds $(z_1, z_u)$ into $ (\xi_1, z)$ by raising them with the private key $\lambda$ under the standard diversion technique~\cite{okamoto_divertible_1990}. The converted proof is eventually transformed to a signature with the Fiat-Shamir technique. Next, the issuer stores $\xi^{\upsilon}$ as the identity of this session. Clearly, $\xi^{\upsilon}$ is easy to map to known $\xi$ which is verified in key generation step. Finally, the user outputs a $cred_u$ with $\Sigma$, say $\Sigma = (\zeta_1,\rho,{\overline{\omega}},\sigma_1,\sigma_2,\delta,m)$ is the signature for the message $m$.

\paragraph{Credential Verification} Credential verification, proceeding after credential generation, is another interactive protocol that runs independently from blockchain involving only the user and the verifier. We say a credential $(\Sigma, m)$ is {\em valid} if it satisfies:
$$
\overline{\omega} + \delta  = \mathcal{H}_{2}( \zeta_1 | g^{\rho}y^{\overline{\omega}} | g^{\sigma_1}\zeta_1^{\delta} | h^{\sigma_2}(z / \zeta_1)^{\delta} | m).
$$

\subsection{Auditable Revocation} 

\paragraph{Credential Tracing} Credential Tracing is an interactive protocol that involves the tracer, the TEE-powered blockchain node and blockchain consensus nodes. It covers the following sub-protocols:

\begin{enumerate}

\item A tracer first fetches the $\pk_{cid}$ of the tracing contract $\widehat{contract}$, and then encrypts the input of the user's identity $\xi^{\upsilon}$ as $inpt_c = \mathcal{ASM}.\mathsf{Enc}(\pk_{cid},\xi^{\upsilon})$ and sends the $\widehat{contract}$ within $inpt_c$ to a TEE-powered blockchain node. Obviously, the input of this smart contract remains secret due to encryption.

\item To start the process of the execution, the TEE-powered blockchain node first loads the contract $\widehat{contract}$, the input $inpt_c$ and the previous encrypted state $state_{init}$ into the contract TEE.

\item The contract TEE decrypts $inp_c$ and $state_{init}$ with the keys from the key manager TEE, and starts to execute the anonymity tracing function with output $I_{cred}$ and state $state_{t}$. Observes that,

\begin{equation}
I_{cred} = (\xi^{\upsilon})^{x_t} = g^{\gamma\upsilon x_t}  = y_t^{\gamma\upsilon} = \zeta_1.
\label{con:credential}
\end{equation}

\item The contract TEE obtains a fresh symmetric-key $\key_{cid}^{out}$ from the key manager TEE and calculates a new encrypted output $outp_{new}^{TEE} = \mathcal{SM}.\mathsf{Enc}(\sk_{cid}^{out},I_{cred})$ and a new encrypted state $state_{new}^{TEE} = \mathcal{SM}.\mathsf{Enc}(\key_{state},state_{t})$. Then, it sends $state_{new}^{TEE}$, $outp_{new}^{TEE}$ and the proper
attestation to the tracer through a secure channel established by the tracer's session keys $(\iota,\tau)$.

\item The tracer acknowledges the reception by calling back the TEE-powered blockchain node, which triggers the contract TEE to send the transaction $tran = (\widehat{contract}, outp_{new}^{TEE},\allowbreak state_{new}^{TEE}, proof)$ to the blockchain. $proof$ is used to protect the integrity of the transaction and the correnctness of the $outp_{new}^{TEE}$ and $state_{new}^{TEE}$.

\item Once the consensus nodes confirm $\widehat{contract}$, the contract TEE decrypts $outp_{new}^{TEE}$ and $state_{new}^{TEE}$ as $outp_{new}^{t}$ and $state_{new}^{t}$ and then sends them to the tracer through the mentioned secure channel.

\item The tracer parses the $outp_{new}^{t}$ and $state_{new}^{t}$ and ultimately learns $I_{cred}$ that is the relationship of the credential and that real owner.  

\end{enumerate}
Among all the sub-protocols, we emphasize that the sub-protocols five and six are atomic operations, and we refer to~\cite{cheng_ekiden:_2018} for more details. Also, we highlight two main features. Firstly, $\widehat{contract}$ will be confirmed by the consensus nodes mentioned in sub-protocol six. Thus, the contract invoked eventually becomes immutable and auditable.
Second, the output $outp_{new}^{t}$ and state $state_{new}^{t}$ are kept secret in the whole life of the execution and transmission.

\paragraph{Identity Tracing} Identity tracing and the credential tracing have the same tracing mechanism. Due to the space limit, we skip its full description. Observes that,
\begin{equation}
I_{id} = \zeta_1^{1/x_t} = z_1^{\lambda/x_t} = y_t^{\upsilon\lambda/x_t} = g ^{\upsilon\lambda} = \xi^{\upsilon}.
\label{con:identity}
\end{equation}

Since $\xi^{\upsilon}$ is stored or published by the issuer, the tracer can instantly identify the user who issued the credential. 

\paragraph{Tracing Inspection} The tracing activities checking is straightforward.  Given the inspector type (identity tracing or credential tracing) and the smart contract identifier, the inspector scans the blockchain to collect all the transactions related to this contract. Then, the inspector checks all these transactions to recognise suspicious activities. 

\section{Implementation and Evaluation}
\label{sec:implementation}

We have implemented a proof of concept of our instantiation. Next we report on our proof of concept and its performance. The corresponding code has been made available open source and is to be found at \url{https://github.com/typex-1/auditable-credential-core}.

\subsection{Implementation}

We focus on implementing the blind issuance protocols and the anonymity revocation smart contracts, and leave the implementation of the menthoned TEE-related protocols to the Oasis Devnet~\cite{cheng_ekiden:_2018}. Specifically, our implementation is divided into two modules: the issuing module and the tracing module. The issuing module covers the protocol of credential generation and credential verification, and it is realised by Python in 168 lines of code. The issuing module is responsible for blindly issuing credentials and verifying the issued ones. Meanwhile, the tracing module which performs the protocol of credential tracing and identity tracing is achieved by Solidity in 449 lines of code and deployed in Oasis Devnet. The tracing module allows the tracer to uncover the identity of a credential or the credential of a specific user. 

\begin{verbatim}
    // example code;
    mapping (address => uint256) private CredentialTraceResults;
    function CredentialRevocation (uint256 upsilon) {
        CredentialTraceResults[upsilon] = power(upsilon, xt, p);
    }
\end{verbatim}

\begin{figure}[ht!]
    \centering
    \includegraphics[width=12cm]{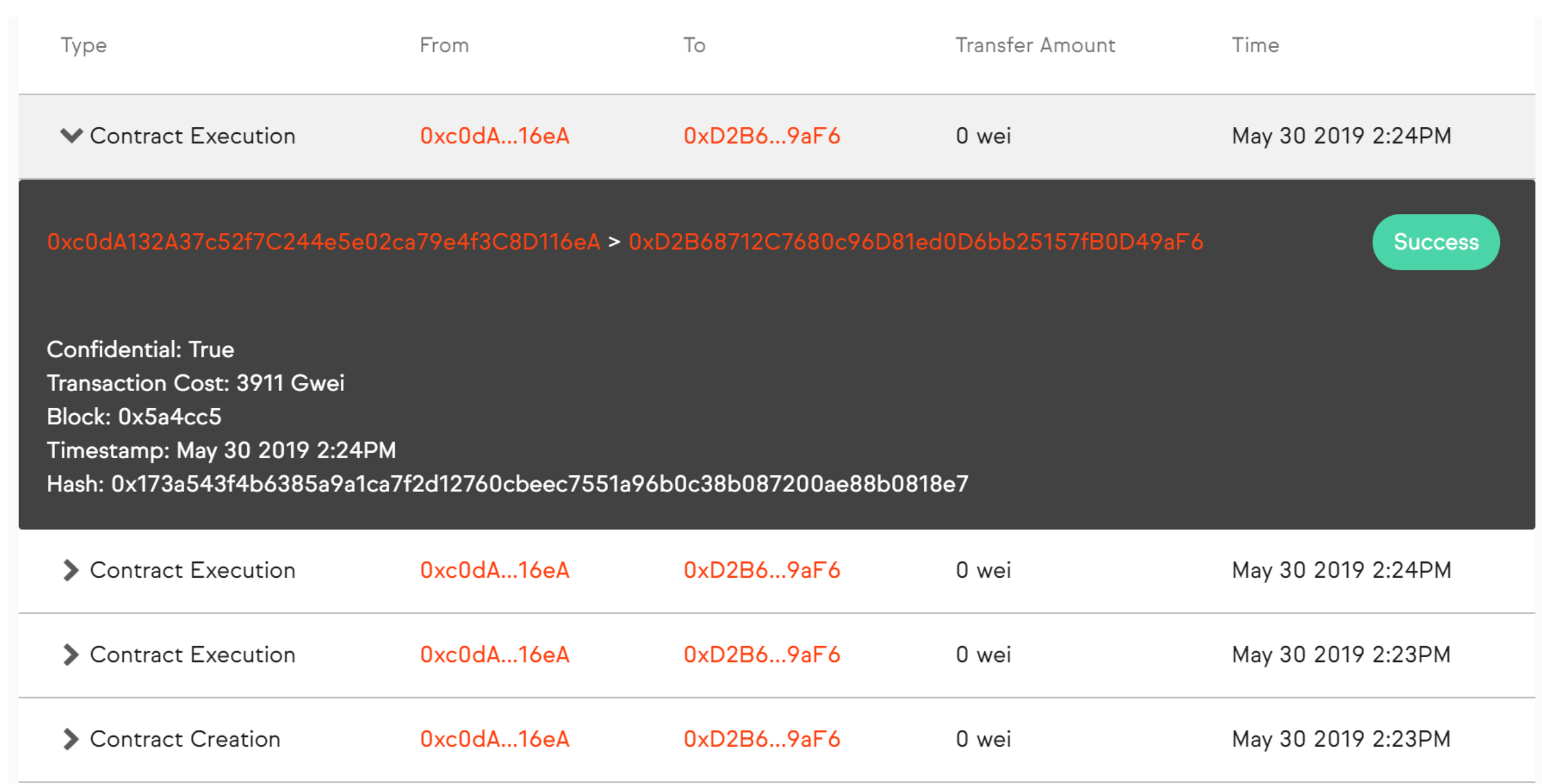}
    \caption{Credential anonymity revocation records.}
    \label{fig:modules}
\end{figure}

Two key properties are highlighted in our implementation: the full protection of private state and the auditable anonymity tracing records. The full protection of private state is represented as that the input data and the output data in the contract are kept secret in the full life cycle. For example, as is shown in the example code, the parameter of \textit{CredentialTraceResults} is designed to privately store the relationship of the identity and credential. The other entities can not read them unless through an end-to-end secure channel that has been established with the contract TEE. The auditability of anonymity tracing records is evident in that all the smart contract invoking records are publicly visible and immutable (The Fig.~\ref{fig:modules} is a smart contract creation and invoking example). In addition, we provide a web-based client to present an interactive process of credential and identity tracing and show the full code in the repository\footnote{https://github.com/typex-1/auditable-credential-core}.

\subsection{Evaluation}
Our performance evaluation covers five operations:  tracing parameter generation,  credential issuing,  credential verifying,  credential tracing and  identity tracing (see Table~\ref{tab:cost}). All experiments are conducted on a Dell precision 3630 tower with 16GB of RAM and one 3.7GHz six core i7-8700K processors running Ubuntu 18.04. Experiments are measured in seconds through wall clock run time, where a time difference is obtained between the start and end of the code execution. To have an accurate and fair test result, we repeat the measure for each execution  300 times and calculate its average. Also, to simplify the performance evaluation, we measure the running time of each step and accumulate them together if there are many steps involved. It is noted that, all operations take much less than one second to complete and the credential issuing is the main performance bottleneck. This operation takes more time than others because issuing a new credential requires many interactions between users and issuers. Fortunately, this bottleneck can be ignored in real applications because after all it meets the nature of credential using scenario, which means a credential is issued only once but could be verified or traced multiple times.

We then examine the operating cost. Similar to the performance testing, the cost evaluation covers five mentioned credential operations. Table~\ref{tab:cost} shows the data size and the cost of these operations in gas under an elliptic curve with 128 bits security level. An analysis of the data size and cost points to some trends. The data size of the operation of parameter generation is the largest since this operation needs to register the group parameters to the smart contract. Surprisingly, the cost of the parameter generation is not the largest as this operation does not cover the complex computations. On the contrary, the data size of the operation of the credential issuing and verifying is zero, and there is no gas cost since these operations are executed independently from the blockchain.  Meanwhile, the credential tracing and the identity tracing have static gas cost since the length of input data of these operations is constant, and the data handling procedure is fixed. In our scheme, a one-time elliptic-curve exponentiation (see Equations~\eqref{con:credential} and~\eqref{con:identity}) is adequate
to conduct the complete tracing activity. The gas cost of the one-time computation is quite lower and more easier to adopt by users when compared with some blockchain-based applications such as~\cite{bunz2019zether,sonnino2018coconut}, where they have massive elliptic-curve exponentiation operations and significant cost.


\begin{table}[htb!]
 \caption{The performance, input data size, cost and latency of various operations.} 
  \centering
  \begin{tabular}{lcccc}
    \toprule
    Operation  & Performance (seconds)  & Size (bytes)  & Gas  & Latency (seconds) \\
    \midrule
    Parameter generation & 0.00084   & 260  & 20672 & 14.781 \\
    Credential issuing  & 0.00740 & 0  & 0  & 1.601 \\
    Credential verifying  & 0.00232 & 0 & 0  &  1.175 \\
    Credential tracing  & 0.00306 & 132 & 390261  & 17.538  \\
    Identity tracing  & 0.00455 & 132 & 388944  & 18.905 \\
    \bottomrule
  \end{tabular}
  \label{tab:cost}
\end{table}

Finally, we conduct latency testing as latency is an essential consideration for adopting a system. For our implementation, the latency time includes blockchain confirming time, the network request time and network response time. It is observed that the latency of credential issuing as well as identity verifying is much smaller than other operations. The main reason behind this is that these two operations run independently from blockchain and do not wait for the block to be confirmed. Meanwhile, the average latency of credential tracing and identity tracing is approximately eighteen seconds, which would be a primary drawback of our system. Given these latency constraints, our system, at least built on the current version of Oasis Devnet is not suitable for applications that require fast credential tracing or identity tracing. However, for some privacy-priority applications such as medical record tracing system, our scheme provides a powerful framework to protect patients' privacy. 


The main roadblock to the business adoption of blockchain is its low throughput of on-chain transaction. Our system, armed with the blockchain and trusted execution environment, suffers the same scalability issues. Fortunately, the flexible smart contract makes our scheme easier to support batch anonymity revealing. This means a tracer can collect a group of credentials and send them the blockchain once. 
With such a mechanism, a massive chunk of tracing transactions can be off-loaded from the blockchain, which mitigates the scalability flaw. Meanwhile, some efforts~\cite{eyal2016bitcoin,zamfir2017casper} have been made to increase the scalability of the blockchain. Our system would benefit from these works.

\section{Example Applications}
\label{sec:application}
Our scheme has numerous practical applications in some privacy-sensitive scenarios. Two typical use cases are described as follows: 
\paragraph{Medical record protection}Our scheme may be used for privacy medical record protection specifically for unrestricted research purposes. A medical record is supposed to be very sensitive in some cases such as in HIV and sexually transmitted infections. A hospital might share the medical record with a research institution without patients' permission thereby causing information leakage. Our mechanism allows the hospital to show the real patient records without knowing the patients' real identities, so  privacy is respected. In the case of family genetic disorder, patients may disclose their identities to the research institution on their own free will by invoking the privacy-preserving smart contract.
\paragraph{Vehicle registration management}Traditionally, the vehicle registration office issues a vehicle plate number knowing all the identities and corresponding car information. If someone with an intent to identify the specific driver colludes with the registration office, privacy invasion occurs. Moreover, the plate number may become a surveillance tool in conjunction with a closed-circuit television camera, which is in wide use almost anywhere. Our scheme allows the issuance of a vehicle plate number without the vehicle registration office knowing the relationship between the number and the driver's identity. Furthermore, it allows the certificate to be traced in an auditable way when some emergencies such as traffic accidents.

\section{Conclusion}
\label{sec:conclusion}

Anonymity credentials and anonymity revocation were proposed several decades ago, but they have not yet gained significant adoption.  Some potential obstacles  are the lack of auditability and neutrality for the revocation process. In this paper, we proposed a blockchain-powered traceable anonymous credential framework. Our approach allows the issuer to blindly issue a credential, then leverages a privacy-preserving smart contract that acts as a revelator to trace the credential. More importantly, all these tracing activities are auditable due to the immutable smart contract calling records provided by the public ledger. The auditability and neutrality guaranteed by the blockchain avoid misuse of tracing and potential collision problems to a great extent.

\noindent \textbf{Future work.} Even if our scheme provides a powerful approach to trace the anonymity with auditability, in practice, it is still possible  that one of a tracer's private keys is stolen or misused. Fortunately, the flexibility of smart contract makes our scheme amenable to support threshold-revealing using well-known multiparty computation techniques.  


\section{Acknowledgement}
The authors would like to thank Feng Liu, Geyang Wang and Alphea Pagalaran for their constructive suggestions on the manuscript. The authors would also like to thank the anonymous referees for their valuable comments that improved the quality of the paper.

\bibliographystyle{splncs04.bst}
\bibliography{reference}

\end{document}